\documentclass[two column]{article}
\usepackage{graphicx}
\usepackage{amsfonts}
\UseRawInputEncoding
\usepackage{amsmath}
\newtheorem{proposition}{Proposition}
\newtheorem{corollary}{Corollary}
\begin{document}

\title{Matched entanglement witness criteria for continuous variables}


\author{Xiao-yu Chen, Maoke Miao, Rui Yin, Jiantao Yuan\\
{\small {School of Information and Electrical Engineering, Zhejiang University City College, Hangzhou {\rm 310015}, China }}}

\date{}


\maketitle

{\it{Abstract}}.-
We use quantum entanglement witnesses derived from Gaussian operators to study the separable criteria of continuous variable states. We transform the validity of a Gaussian witness to a Bosonic Gaussian channel problem. It follows that the maximal means of two-mode and some four-mode Gaussian operators over product pure states are achieved by vacuum (or coherent states and squeezed states) according to the properties of Bosonic Gaussian channels. Then we have necessary and sufficient criteria of separability not only for Gaussian quantum states, but also for non-Gaussian states prepared by photon adding to or/and subtracting from Gaussian states. The criterion can be further explicitly expressed with covariance matrix of the Gaussian state or covariance matrix of Gaussian kernel of the non-Gaussian state. This opens a way for precise detection of non-Gaussian entanglement.



{\it {Introduction}} -
Quantum entanglement is an indispensable resource for quantum information processing. For a general quantum state, it is by no means easy to decide whether it is entangled or not. Many entanglement criteria have been proposed to detect entanglement \cite{HorodeckiR}\cite{Guhne}. Usually the criteria are first proposed for qubit systems, then extended to continuous variable (CV) systems. Among them is positive partial transpose (PPT) criterion \cite{Peres}\cite{Horodecki}. The applications of PPT criterion for CV include separable conditions for two-mode Gaussian states \cite{Simon}, multi-mode bipartite Gaussian states \cite{Werner}\cite{Wang} and general CV states \cite{Shchukin}\cite{Chen07}\cite{ZhangDa}, respectively. The fail of PPT criterion is the existence of the bounded entangled state. Bound entangled Gaussian states were found in \cite{Werner}, and algorithms have been developed to find the bound entangled states in \cite{Giedke}. The applications of the other criteria to CV systems are, uncertainty principle criterion for Gaussian states\cite{Duan} and non-Gaussian states\cite{Agarwal}\cite{Hillery}\cite{Nha}, computable cross norm and realignment criterion (CCNR) \cite{Jiang14}, entropy criterion \cite{Walborn}, local operator criterion \cite{Zhang} and so on.

Entanglement can also be detected with entanglement witnesses \cite{Guhne}\cite{SperlingVogel}\cite{Huber}. An entanglement witness is a Hermitian operator which has non-negative means on all separable states and has a negative mean at least on one entangled state. The definition of the entanglement witness leads to an inequality on the witness operator. The optimization of the inequality fixes one of the parameters of the witness, it is called weakly optimal \cite{Guhne}. A full optimization \cite{Chen17}\cite{ChenPRA20}\cite{ChenJSAC} of all the parameters of a witness will lead to a matched entanglement witness. The weakly optimization leads to a necessary criterion of separability, and the full optimization leads to a sufficient criterion of separability. The matched entanglement witness is capable of detecting different types of entanglement in multipartite entanglement \cite{Chen17}.  The full optimization of witness was applied to detect the entanglement structure of a six-mode CV state \cite{Gerke}, with the witness chosen as the quadratic combination of position and momentum operators.

{\it{Criterion from witness}} -
Assuming a witness with the form of $\hat{W}=\Lambda \hat{I}-\hat{M}$, where $\hat{I}$ is the identity operator. The validity of a witness requires that $Tr(\hat{\rho}_{s}\hat{W})=\Lambda-Tr(\hat{\rho}_{s}\hat{M})\geq0$ for all separable states $\hat{\rho}_{s}$. For any given Hermitian operator $\hat{M}$ (named as a detector), let $\Lambda=\max_{\hat{\rho}_{s}}Tr(\hat{\rho}_{s}\hat{M}),$  then $\hat{W}$ is a weakly optimal entanglement witness. If $Tr(\hat{\rho}\hat{W})=\Lambda-Tr(\hat{\rho}\hat{M})<0$ for a state $\hat{\rho}$, then $\hat{\rho}$ is an entangled state. Hence for a given detector $\hat{M}$, there is a necessary criterion of separability $Tr(\hat{\rho}\hat{M})\leq\Lambda.$
 For a given state $\hat{\rho}$, we define $\mathcal{L}=\min_{\hat{M}}\frac{\Lambda}{Tr(\hat{\rho} \hat{M})}$ with $\Lambda>0$ and $Tr(\hat{\rho} \hat{M})>0$. The refined separable criterion is
 \begin{equation}\label{we0}
   \mathcal{L}\geq 1.
 \end{equation}
The extreme detector, $\hat{M^{*}}$, that minimizes $\frac{\Lambda}{Tr(\hat{\rho} \hat{M})}$, leads to a matched entanglement witness, $\hat{W^{*}}=\Lambda \hat{I}-\hat{M^{*}}$. Notice that (\ref{we0}) is a necessary and sufficient criterion for separability. Since if $\mathcal{L}< 1$, then the state $\hat{\rho}$ is entangled. While if $\mathcal{L}\geq 1$, the state $\hat{\rho}$ should be separable. Otherwise, if there is an entangled state $\hat{\rho}$ such that $\mathcal{L}\geq 1$, it means that there is no witness which can detect this entangled state. However, it has been proved that `for each entangled state $\hat{\rho}$, there exists an entanglement witness detecting it' \cite{Guhne}\cite{Horodecki}.

{\it{Gaussian witness}} -
For CV systems, we use the notation $\hat{R}=(\hat{x}_{1},\hat{p}_{1},...,\hat{x}_{n},\hat{p}_{n})^T$ for an operator vector containing the position $\hat{x}_{j}$ and momentum $\hat{p}_{j}$ quadratures of all modes $(j=1,...,n)$. The commutation relations read $[\hat{R}_{j},\hat{R}_{l}]=i\sigma_{jl}$, where the symplectic matrix takes the form $\sigma=\left(
                                                                               \begin{array}{cc}
                                                                                 0 & 1 \\
                                                                                 -1 & 0 \\
                                                                               \end{array}
                                                                             \right)
^{\oplus n}$. The covariance matrix (CM) $\gamma$ of a state $\hat{\rho}$ can be written in terms of $\gamma_{ij}=\frac{1}{2}\langle\hat{R}_{i}\hat{R}_{j}+\hat{R}_{j}\hat{R}_{i}\rangle_{\hat{\rho}}-\langle \hat{R}_{i}\rangle_{\hat{\rho}}\langle\hat{R}_{j}\rangle_{\hat{\rho}}$. A local displacement does not affect the entanglement, which implies that we only need to consider the case of $\langle \hat{R}\rangle_{\hat{\rho}}=0$. The characteristic function of a state is defined as $\chi(z)=Tr[\hat{\rho}\exp(iz\hat{R})]$, where $\exp(iz\hat{R})$ is called Weyl operator. A Gaussian state is completely characterized by its CM, that is, $\chi(z)=\exp{(-\frac{1}{2}z\gamma z^T)}$ when the first moment is set to zero.
Sometimes, it is more convenient to deal with the annihilation operator and creation operator. Let the Weyl operator $\exp(iz\hat{R})$ be equal to the displacement operator $D(\mu)=\exp[\Sigma_{j=1}^{n}(\mu_j\hat{a}_{j}^{\dagger}-\mu_{j}^*\hat{a}_{j})]$, where $\hat{a}_{j}=\frac{1}{\sqrt{2}}(\hat{x}_{j}+i\hat{p}_{j})$ and $\hat{a}_{j}^{\dagger}=\frac{1}{\sqrt{2}}(\hat{x}_{j}-i\hat{p}_{j})$ are the annihilation and creation operators of the $j$-th mode, respectively. Then, we have $\mu_j=\frac{1}{\sqrt{2}}(-z_{2j}+iz_{2j-1})$. The characteristic function, $\chi(z)=\chi(\mu)=\exp[-\frac{1}{2}(\mu,\mu^*)\tilde{\gamma}(\mu,\mu^*)^T]$, with complex covariance matrix (CCM) $\tilde{\gamma}$ is a transform of the CM $\gamma$.

 We choose a zero-mean Gaussian operator $\hat{M}$ as the detector. We call the corresponding witness operator $\hat{W}$ Gaussian witness for simplicity. For the validity of the witness, we need to calculate the maximal mean of the detector $\hat{M}$ over product states. We have $\Lambda=\max_{\hat{\rho}_{A},\hat{\rho}_{B}}Tr[\hat{M}(\hat{\rho}_{A}\otimes\hat{\rho}_{B})]$, if the system is divided into A and B subsystems. It suffices to consider $\hat{\rho}_{A}$ and $\hat{\rho}_{B}$ to be pure states, so $\Lambda=\max_{\psi_{A},\psi_{B}}\langle \psi_A|\langle \psi_B|\hat{M}|\psi_A\rangle|\psi_B\rangle$.

 We deal with $\langle\psi_B|\hat{M}|\psi_B\rangle$ as the output matrix of a map $\hat{M}$ when the input is the pure state $|\psi_B\rangle$. The output matrix then has a largest eigenvalue $\Lambda_{|\psi_B\rangle}$ with corresponding eigenvector $|\phi\rangle$. Let $|\psi_{A}\rangle=|\phi\rangle$, our problem of maximizing the mean of $\hat{M}$ over the product pure state reduces to the problem of maximizing the largest eigenvalue $\Lambda_{|\psi_B\rangle}$ of the output matrix of map $\hat{M}$ with respect to input state $|\psi_B\rangle$. If the map $\hat{M}$ represents one of the four quantum Bosonic Gaussian channels\cite{Holevo}\cite{Caruso}\cite{GiovannettiNP}, then it is known that vacuum state as the input will minimize the output entropy \cite{GiovannettiNP} \cite{GiovannettiCM}, furthermore it will maximize the output majorization\cite{GiovannettiNC}\cite{GiovannettiTM}. For the majorization of a state $\rho$, we say $\rho_{2}$ majorizes $\rho_{1}$ if
 \begin{equation}\label{we1}
   \sum_{j=1}^{k}\lambda_{j}^{\rho_{1}}\leq\sum_{j=1}^{k}\lambda_{j}^{\rho_{2}},   \forall k\geq1,
 \end{equation}
where $\lambda_{j}^{\rho} (j=1,...,k)$ are the eigenvalues of $\rho$ in descending order. Clearly, if $\rho_{2}$ majorizes $\rho_{1}$, then the largest eigenvalue of $\rho_{2}$ is not less than that of $\rho_{1}$. Hence, vacuum state input maximizes the largest eigenvalue of the Bosonic Gaussian channel output. As far as we prove that the map $\hat{M}$ represents a Bosonic Gaussian channel, the maximal mean of $\hat{M}$ over product states will be shown to be achieved by vacuum (or coherent states and squeezed states). In the following, we will prove that the $1\times1$ (one mode at each party) and some $2\times2$ (with two modes at both parties)  Gaussian detectors are Bosonic Gaussian channels.

{\it{Two mode Gaussian state}} -
For a $1\times1$ system, consider a detector $\hat{M}$ with standard form of CM, $\gamma_{M}$, without loss of generality. The CM has the form of
\begin{equation}\label{we2}
  \gamma_M=\left(
           \begin{array}{cccc}
             \mathcal{M_A} & \mathcal{M_C} \\
             \mathcal{M_C} & \mathcal{M_B} \\
           \end{array}
         \right),
\end{equation}
where $\mathcal{M_A}=diag(M_{1},M_{1}), \mathcal{M_B}=diag(M_{3},M_{3}), \mathcal{M_C}=diag(M_{5},-M_{6})$.

Let $|\psi_B\rangle=\sum_{k}a_{k}|k\rangle$, then the application of map $\hat{M}$ on the basis $|k\rangle\langle m|$ leads to output element $(|k\rangle\langle m|)_{out}=Tr_{B}(\hat{M}|k\rangle\langle m|)$. Let the characteristic function of the basis $|k\rangle\langle m|$ and $(|k\rangle\langle m|)_{out}$ be $\chi_{in}(|k\rangle\langle m|,\mu)$ and $\chi_{out}(|k\rangle\langle m|,\mu)$, respectively. Then
 \begin{equation}\label{we2a}
  \chi_{in}(|k\rangle\langle m|,\mu)=\langle m|D(\mu)|k\rangle=\frac{(-1)^k}{\sqrt{m!k!}}e^{-\frac{|\mu|^2}{2}}H_{mk}(\mu,\mu^{*})
 \end{equation}
 with Hermitian polynomial
 \begin{eqnarray*}
  H_{mk}(\mu,\mu^{*})=\sum_{l=0}^{\min(m,k)}\frac{m!k!}{(m-l)!(k-l)!l!}(-1)^{l}\mu^{m-l}\mu^{*k-l}\nonumber\\
  =\frac{\partial^{m+k}}{\partial t^m\partial t'^k} \exp[-tt'+\mu t+\mu^{*}t']|_{t=t'=0}.
 \end{eqnarray*}
Denote $\nu=(\nu_{1},\nu_{2})$, the characteristic function of $\hat{M}$ is $\chi_{M}(\nu)$, then $(|k\rangle\langle m|)_{out}$ is
\begin{eqnarray}\label{we2b}
  && Tr\int\chi_{M}(\nu)D(-\nu)\chi_{in}(|k\rangle\langle m|,\mu)D(-\mu)[\frac{d^2\nu}{\pi}] \frac{d^2\mu}{\pi}\nonumber\\
  &&=\int\chi_{M}(\nu)D(-\nu_{1})\chi_{in}(|k\rangle\langle m|,-\nu_{2})[\frac{d^2\nu}{\pi}] \nonumber\\
\end{eqnarray}
The integral on $\nu_{2}$ can be carried out by using (\ref{we2a}) and interchanging the order of integral and partial derivatives. The characteristic function of the output basis element $(|k\rangle\langle m|)_{out}$ is
\begin{eqnarray}\label{we2c}
  \chi_{out}(|k\rangle\langle m|,\nu_{1})=\exp[-M_{1}|\nu_{1}|^2+\frac{1}{M_{3}'}|\tau|^2]\frac{(-1)^k}{M_{3}'\sqrt{m!k!}}\nonumber \\
  \times\frac{\partial^{m+k}}{\partial t^m\partial t'^k}\exp[-tt'(1-\frac{1}{M_{3}'})-\frac{\tau^{*}t}{M_{3}'}-\frac{\tau t'}{M_{3}'}]|_{t=t'=0},
\end{eqnarray}
where $\tau=\frac{1}{2}[(M_{5}+M_{6})\nu_{1}+(M_{5}-M_{6})\nu_{1}^{*}]$, $M_{3}'=M_{3}+\frac{1}{2}$. Let $s=\sqrt{1-\frac{1}{M_{3}'}}t, s'=\sqrt{1-\frac{1}{M_{3}'}}t'$, then we can put the characteristic function in the following form
\begin{eqnarray}\label{we2d}
  \chi_{out}(|k\rangle\langle m|,\nu_{1})=\frac{(-1)^k\exp[-M_{1}|\nu_{1}|^2+\frac{1}{M_{3}'}|\tau|^2]}{M_{3}'(1-\frac{1}{M_{3}'})^{(m+k)/2}\sqrt{m!k!}}\nonumber\\
  \times H_{mk}(-\frac{\tau^*}{\sqrt{M_{3}'(M_{3}'-1)}},-\frac{\tau}{\sqrt{M_{3}'(M_{3}'-1)}})\nonumber\\
  =\frac{1}{M_{3}'(1-\frac{1}{M_{3}'})^{(m+k)/2}}\chi_{in}(|k\rangle\langle m|,-\frac{\tau^*}{\sqrt{M_{3}'(M_{3}'-1)}})\nonumber\\
  \times\exp[-M_{1}|\nu_{1}|^2+\frac{M_{3}|\tau|^2}{M_{3}'(M_3'-1)}].
\end{eqnarray}
The definition of a Bosonic Gaussian channel is that its characteristic function undergoes a transformation of \cite{Holevo}
\begin{equation}\label{we2d1}
  \chi_{out}(z)=\chi_{in}(Kz)e^{-\frac{1}{2}z\alpha z^T},
\end{equation}
where the real vector, $z=(z_{1},z_{2})$, is related to our complex variable $\nu_{1}$ through $\nu_{1}=\frac{1}{\sqrt{2}}(-z_{2}+iz_{1})$. Here, $K$ is a linear transformation in symplectic space and $\alpha$ is a Hermitian matrix. For one mode channel, when $\det K>0 / \det K<0$, it can always be transformed to gauge covariant/contravariant channel by proper symplectic transformations of input and output states\cite{Caruso}. Clearly, when $M_{3}$ tends to infinitive, the factor $(1-\frac{1}{M_{3}'})$ can be removed from (\ref{we2d}), the map $\hat{M}$ then is a Bosonic Gaussian channel. The complete positivity of the channel will lead to an inequality on $\alpha$ and $K$. This inequality turns out to be that $\hat{M}$ is a Gaussian quantum state for gauge covariant channel, and $\hat{M}$ is a separable Gaussian quantum state for contravariant channel.

Next, we will consider a more general Gaussian detector $\hat{M}$ with 6 parameter CM. The CM has still the form of (\ref{we2}), but with $\mathcal{M_A}=diag(M_{1},M_{2}), \mathcal{M_B}=diag(M_{3},M_{4}), \mathcal{M_C}=diag(M_{5},-M_{6})$ instead. With a local squeezing operation $U_{A}\otimes U_{B}$, the detector $\hat{M}$ can be transformed to another detector $\hat{M}_{s}=U_{A}\otimes U_{B}\hat{M}U_{A}^\dag\otimes U_{B}^\dag$ with a standard form of CM. We have $Tr(M\rho_{A}\otimes\rho_{B})=Tr(M_{s}U_{A}\rho_{A}U_{A}^\dag \otimes U_{B}\rho_{B}U_{B}^\dag)$. So that the 6 parameter detector $\hat{M}$ is a Bosonic Gaussian channel if its parameters tend to infinitive. It follows that, its maximal mean over pure product states is
\begin{equation}\label{we2e}
  \Lambda=\frac{1}{\min_{\gamma_{A},\gamma_{B}}\sqrt{\det(\gamma_{M}+\gamma_{A}\oplus\gamma_{B})}},
\end{equation}
where $\gamma_{A}=\frac{1}{2}diag(x,1/x),\gamma_{B}=\frac{1}{2}diag(y,1/y)$ are the CMs of squeezed vacuum states of the two subsystems.

 We then have the following proposition from the definition of the witness.
\begin{proposition}\label{proposition1}
For a $1\times1$  Gaussian state with CM $\gamma$, the necessary and sufficient criterion of separability is
\begin{equation}\label{we2f}
   \mathcal{L}^2=\min_{\hat{M}}\frac{\det(\gamma+\gamma_{M})}{\min_{\gamma_{A},\gamma_{B}}\det(\gamma_{A}\oplus\gamma_{B}+\gamma_{M})}\geq 1,
\end{equation}
where $\gamma_{M}\geq\frac{i}{2}\sigma$ for $\hat{M}$ is a Gaussian state, and $\det(\gamma_{A})=\det(\gamma_{B})=\frac{1}{4}$ for pure Gaussian subsystem states.
\end{proposition}
This is the Gaussian witness entanglement criterion for a two-mode Gaussian state. Notice that the necessary and sufficient criterion for a two-mode Gaussian state is well known\cite{Simon}\cite{Duan}, we should check if Proposition \ref{proposition1} lead to the same explicit condition of separability.

The standard CM of a two-mode Gaussian state is given by
\begin{equation}\label{we2g}
  \gamma=\left(
           \begin{array}{cccc}
             \mathcal{A} & \mathcal{C} \\
             \mathcal{C} & \mathcal{B} \\
           \end{array}
         \right),
\end{equation}
where $\mathcal{A}=diag(a,a), \mathcal{B}=diag(b,b),  \mathcal{C}=diag(c_{1},-c_{2})$.  We use the 6 parameter $\gamma_{M}$ described above with diagonal elements $M_{1},M_{2},M_{3},M_{4}$ and off-diagonal elements $M_{5},-M_{6}$. Then
\begin{equation}\label{we2h}
\mathcal{L}^2=\min_{\hat{M}}\max_{x,y}f_{1}(M_{1},M_{3},M_{5},x,y)f_{2}(M_{2},M_{4},M_{6},x,y),
\end{equation}
with
\begin{eqnarray*}
  f_{1}(M_{1},M_{3},M_{5},x,y)=\frac{(M_{1}+a)(M_{3}+b)-(M_{5}+c_{1})^2}{(M_{1}+\frac{x}{2})(M_{3}+\frac{y}{2})-M_{5}^2},\nonumber\\
  f_{2}(M_{2},M_{4},M_{6},x,y)=\frac{(M_{2}+a)(M_{4}+b)-(M_{6}+c_{2})^2}{( M_{2}+\frac{1}{2x})(M_{4}+\frac{1}{2y})-M_{6}^2}.
\end{eqnarray*}
We then change the order of minimization and maximization in (\ref{we2h}). There have $(M_{1}+a)(M_{3}+b)-(M_{5}+c_{1})^2=A_{1}+A_{2}+A_{3}$ with $A_{1}=(M_{1}+\frac{x}{2})(M_{3}+\frac{y}{2})-M_{5}^2$, $A_{2}=(M_{1}+\frac{x}{2})(b-\frac{y}{2})+(M_{3}+\frac{y}{2})(a-\frac{x}{2})-2M_{5}c_{1}$, $A_{3}=(a-\frac{x}{2})(b-\frac{y}{2})-c_{1}^2$. Notice that $A_{2}\geq A_{2}':=2(\sqrt{(M_{1}+\frac{x}{2})(M_{3}+\frac{y}{2})}\sqrt{(a-\frac{x}{2})(b-\frac{y}{2})}-M_{5}c_{1})$, we may use $A_{2}'$ to substitute $A_{2}$ in the minimization of $f_{1}$ with respect $M_{1},M_{3},M_{5}$. For sufficiently large $M_{1},M_{3},M_{5}$, we just omit $A_{3}$ term, then $\min f_{1}=1+\min_{M_{1},M_{3},M_{5}}\frac{A_{2}'}{A_{1}}$. Notice that $f_{1}$ and $f_{2}$ are independent in minimizations, we should keep both $\min f_{1}\geq1$ and $\min f_{2}\geq1$ to preserve (\ref{we2h}). We should keep $A_{2}'\geq0$, which is only possible when $\sqrt{(a-\frac{x}{2})(b-\frac{y}{2})}-c_{1}\geq 0$. Otherwise, we can make $M_{5}$ approach $\sqrt{(M_{1}+\frac{x}{2})(M_{3}+\frac{y}{2})}$ to violate $A_{2}'\geq0$. Thus, we have the following two conditions derived from (\ref{we2f}).
\begin{eqnarray}
  (a-\frac{x}{2})(b-\frac{y}{2})-c_{1}^2\geq0 \label{we2j},\\
  (a-\frac{1}{2x})(b-\frac{1}{2y})-c_{2}^2\geq0 \label{we2k}.
\end{eqnarray}
The equalities in (\ref{we2j}) and (\ref{we2k}) are drawn as two curves in $(x,y)$ plane. If the two curves intersect with each other, we have solution for the two inequalities. The combination of the two equalities in (\ref{we2j}) and (\ref{we2k}) leads to a quadratic equation for $x$ (or $y$). The condition for the existence of the real $x$ solution can be obtained easily. So the existence of the solution $(x,y)$ for both inequalities (\ref{we2j}) and (\ref{we2k}) leads to the following condition
\begin{equation}\label{we2l}
  (ab-c_{1}^2)(ab-c_{2}^2)-\frac{1}{2}|c_{1}c_{2}|-\frac{1}{4}(a^2+b^2)+\frac{1}{16}\geq0.
\end{equation}
It is just the necessary and sufficient separable condition derived by Simon \cite{Simon} with PPT criterion. Notice that the existence of even a point $(x,y)$ as the solution of inequalities (\ref{we2j}) and (\ref{we2k}) means that we have a product squeezed state specified by $x$ and $y$ at hand, the two-mode Gaussian state with CM $\gamma$ can be generated from this product squeezed state by applying displacement operations with classical probability distribution. So, inequalities (\ref{we2j}) and (\ref{we2k}) are also sufficient conditions for separability. We have shown that $\mathcal{L}^2\geq1$ leads to condition (\ref{we2l}). On the other hand, if the condition (\ref{we2l}) is fulfilled, the two inequalities (\ref{we2j}) and (\ref{we2k}) are true for point $(x,y)$ in some range of the plane. Then, we have $\mathcal{L}^2\geq1$. Hence, the Proposition \ref{proposition1} is equivalent to the result of PPT criterion for a two-mode Gaussian state. Thus, the method of entanglement witness is feasible in deriving exact separable criterion for CV systems.

{\it{Bound entangled Gaussian state}} -
The CM of Werner-Wolf bounded entangled $2\times2$ state \cite{Werner} can be extended to a more general form of (The state then is called generalized Werner-Wolf state)
\begin{equation}\label{we3}
 \gamma =\left(
    \begin{array}{cccccccc}
      A & 0 & 0 & 0 & E & 0 & 0 & 0 \\
      0 & B & 0 & 0 & 0 & 0 & 0 & -F \\
      0 & 0 & A & 0 & 0 & 0 & -E & 0 \\
      0 & 0 & 0 & B & 0 & -F & 0 & 0 \\
      E & 0 & 0 & 0& C & 0 & 0 & 0 \\
      0 & 0 & 0 & -F & 0 & D & 0 & 0 \\
      0 & 0 & -E & 0 & 0 & 0 & C & 0 \\
      0 & -F & 0 & 0 & 0 & 0 & 0 & D \\
    \end{array}
  \right).
\end{equation}
 The system is divided as the first two modes for Alice versus the last two modes for Bob. Proper local squeezing operations will transform $\gamma$ to a standard form CM with 4 parameters. For the 4 parameter CM state, we consider a standard $\gamma_{Ms}$, which has the same structure of $\gamma$ with ($A,B,C,D,E,F$) substituted by ($M_{1},M_{1},M_{3},M_{3},M_{5},M_{6}$), respectively. It is convenient to work with CCM in proving $\hat{M}$ to be a Bosonic Gaussian channel. The CCM, $\tilde{\gamma}_{Ms}$, of $\hat{M}$ is
\begin{equation*}\label{we3a}
\left(
    \begin{array}{cccccccc}
      0 & 0 &-N_5 &N_6 & M_1 & 0 & N_5 &N_6 \\
      0 & 0 &N_6 & N_5 & 0 & M_1 &N_6 &-N_5\\
      -N_5 &N_6 & 0 & 0 & N_5 &N_6 & M_3 & 0 \\
      N_6 &  N_5 & 0 & 0 &N_6 &-N_5 & 0 & M_3\\
      M_1 & 0 & N_5 & N_6& 0 & 0 & -N_5& N_6 \\
      0 & M_1 & N_6 &-N_5 & 0 & 0 & N_6 &N_5 \\
      N_5 & N_6& M_3 & 0 & -N_5 & N_6& 0 & 0 \\
      N_6 & -N_5 & 0 & M_3 & N_6 & N_5 & 0 & 0 \\
    \end{array}
  \right),
\end{equation*}
with $N_{5}=M_{5}/2, N_{6}=-M_{6}/2$. The characteristic function of $\hat{M}$ will be $\chi(\mu)=\exp[-\frac{1}{2}(\mu,\mu^*)\tilde{\gamma}_{Ms}(\mu,\mu^*)^T]$, with $\mu=(\mu_{1},\mu_{2},\mu_{3},\mu_{4})$.

With the similar process as in $1\times1$ case, we obtain the output characteristic function $\chi_{out}(\mu_{1},\mu_{2})$ when the input is a pure state at Bob's hand with characteristic function $\chi_{in}(\mu_{3},\mu_{4})$ in the case of infinitive $M_{3}$. We have
\begin{eqnarray}\label{we3b}
&\chi_{out}(\mu_{1},\mu_{2})=\frac{1}{M_{3}'^2}\exp[-(M_1-\frac{M_{6}^2M_{3}}{M_{3}'(M_{3}'-1)})(\mu_{1R}^2+\mu_{2R}^2)]\nonumber\\
&\times \exp[-(M_1-\frac{M_{5}^2M_{3}}{M_{3}'(M_{3}'-1)})(\mu_{1I}^2+\mu_{2I}^2)]\nonumber\\
&\times \chi_{in}(-\frac{M_{6}\mu_{2R}-iM_{5}\mu_{1I}}{\sqrt{M_{3}'(M_{3}'-1)}},-\frac{M_{6}\mu_{1R}+iM_{5}\mu_{2I}}{\sqrt{M_{3}'(M_{3}'-1)}}),
\end{eqnarray}
where $\mu_{jR},\mu_{jI}$ are the real and imaginary parts of $\mu_{j}$. Thus $\hat{M}$ is a Bosonic Gaussian channel\cite{Caruso}\cite{HolevoWerner}. A critical point is that we should prove that this channel is equivalent to tensor product of two-one mode channels. That is, the matrices $K$ and $\alpha$ of the channel should be simultaneously diagonalizable \cite{GiovannettiNC}. By comparing (\ref{we3b}) with (\ref{we2d1}), we have
\begin{equation}\label{we3c}
  K= \frac{1}{\sqrt{M_3'(M_3'-1)}}\left(
                                   \begin{array}{cccc}
                                     M_5 & 0 & 0 & 0 \\
                                     0& 0 & 0 & -M_6 \\
                                     0 & 0 & -M_5 & 0 \\
                                     0 & -M_6 & 0 & 0 \\
                                   \end{array}
                                 \right)
\end{equation}
and $\alpha=diag(\alpha_{1},\alpha_{2},\alpha_{1},\alpha_{2}),$ with $\alpha_{1}=M_1-\frac{M_{5}^2M_{3}}{M_{3}'(M_{3}'-1)}, \alpha_{2}=M_1-\frac{M_{6}^2M_{3}}{M_{3}'(M_{3}'-1)}$. So $K$ and $\alpha$ commutate with each other, they can be simultaneously diagonalized. Then $\hat{M}$ represents a tensor product of one-mode gauge covariant (or contracovariant) channels. The theorem that pure Gaussian input maximizes the channel output majorization can be applied\cite{GiovannettiNC}. Hence, for the 4 parameter 4-mode Gaussian detector $\hat{M}$, we have proven that its maximal mean over product pure states is achieved by the product of Gaussian pure states.

 We then consider a 6 parameter detector with CM, $\gamma_{M}$, being the same structure of $\gamma$ in (\ref{we3}), with ($A,B,C,D,E,F$) substituted by ($M_{1},M_{2},M_{3},M_{4},M_{5},M_{6}$), respectively. Then such a detector is also a Bosonic Gaussian channel, because it can be transformed into a detector with 4 parameter standard form of CM using local squeezing. Then, we have the following result. For the new 6 parameter Gaussian detector of $2\times2$ system, the maximal mean of $\hat{M}$ over the product pure states is $\Lambda$, with
\begin{eqnarray}\label{we3d}
  \frac{1}{\Lambda}=\min_{x,y}[(M_{1}+\frac{x}{2})(M_{3}+\frac{y}{2})-M_{5}^2]\nonumber\\
  \times[(M_{2}+\frac{1}{2x})(M_{4}+\frac{1}{2y})-M_{6}^2].
\end{eqnarray}
It follows the necessary and sufficient condition for the separability of generalized Werner-Wolf states
\begin{eqnarray}\label{we3e}
  \mathcal{L}=\min_{\hat{M}}\max_{x,y}\frac{(M_{1}+A)(M_{3}+C)-(M_{5}+E)^2}{(M_{1}+\frac{x}{2})(M_{3}+\frac{y}{2})-M_{5}^2} \nonumber\\
  \times\frac{(M_{2}+B)(M_{4}+D)-(M_{6}+F)^2}{( M_{2}+\frac{1}{2x})(M_{4}+\frac{1}{2y})-M_{6}^2}\geq 1.
\end{eqnarray}
It is very similar to the case of the $1\times1$ Gaussian state. The solution is almost the same. We arrive at the explicit  necessary and sufficient criterion of separability for the generalized Werner-Wolf state as follows.
\begin{equation}\label{we3f}
  (AC-E^2)(BD-F^2)-\frac{1}{2}|EF|-\frac{1}{4}(CD+AB)+\frac{1}{16}\geq 0.
\end{equation}
The sufficiency of the separability criterion (\ref{we3f}) can be seen from the fact that $(A-\frac{x}{2})(C-\frac{y}{2})-E^2\geq0$ and $(B-\frac{1}{2x})(D-\frac{1}{2y})-F^2\geq0$ as the result of (\ref{we3e}). These two inequalities mean that $\gamma$ in (\ref{we3}) is larger than the CM of a four mode product squeezed state specified with $(x,y)$, so the generalized Werner-Wolf state can be generated from the product state with local operations (classical probability distribution of the first moment or displacement), thus, is separable.

Werner and Wolf have constructed the five parameter series of $2\times2$ Gaussian states \cite{Werner} and have shown that these states are bounded entangled with a quite sophisticated way. A direct calculation using (\ref{we3f}) will show that these states are entangled. The five parameters are $a,b,c,d,e>0$ (we abuse $a$ and $b$ here for a while) with $ad-bc>0$ and $ce-a>0 $ \cite{Werner}. Then  $A=\frac{de-b}{2(ce-a)},  B=\frac{a}{2b}, C=\frac{c(da-bc)}{2(ce-a)},  D=\frac{eb+d}{2b(ad-bc)}, E=\frac{ad-bc}{2(ce-a)}, F=\frac{1}{2b}$ from the state description. The left hand side of (\ref{we3f}) is equal to $-\frac{(ad-bc)}{16b(ce-a)}$, which is always negative. Therefore, the states are entangled.

{\it{Gaussian detector for Non-Gaussian entanglement}} -
The entanglement detecting power of the witness constructed from a Gaussian detector $\hat{M}$ discussed so far is limited to Gaussian states. The entanglement of a non-Gaussian state prepared by adding photon to or/and subtracting photon from a Gaussian state (NGPASG) can also be detected with witness based on a Gaussian detector $\hat{M}$. Consider a non-Gaussian state prepared with photon addition or/and subtraction, and the state can be written as $\hat{\rho}=\mathcal{N}\hat{a}^{k\dagger}\hat{a}^{m}\hat{\rho}_{G}\hat{a}^{m\dagger}\hat{a}^{k}$, where $\hat{\rho}_{G}$ is a zero-mean Gaussian state (we call it Gaussian kernel of state $\hat{\rho}$ ) with CM, $\gamma_{G}$ (CCM $\tilde{\gamma}_{G}$), $\mathcal{N}$ is the normalization. Herein $\hat{a}^k$ is a brief notation of $\hat{a}_{1}^{k_{1}}\hat{a}_{2}^{k_{2}}...\hat{a}_{n}^{k_{n}}$ and similar notation for the creation operator. The non-Gaussian state can be generated from Gaussian kernel operator function,
  \begin{equation}\label{we8}
    \hat{Q}(\xi,\eta)=e^{\xi \hat{a}^{\dagger}}e^{-\eta^* \hat{a}}\hat{\rho}_{G}e^{\eta\hat{a}^{\dagger}}e^{-\xi^* \hat{a}},
  \end{equation}
   by derivatives. Namely, $\hat{\rho}=\mathcal{N}\mathcal{O}\hat{Q}(\xi,\eta)$, where $\xi \hat{a}^{\dagger}=\sum_{j=1}^{n}\xi_{j} \hat{a}_{j}^{\dagger}$ and $\mathcal{O}=(-1)^{|k|+|m|}\frac{\partial^{2|k|+2|m|}}{\partial\xi^{k}\partial\xi^{*k}\partial\eta^{m}\partial\eta^{*m}}|_{\xi=\xi^{*}=\eta=\eta^{*}=0}$, with $|k|=\sum_{j=1}^{n}k_{j}$ and $\partial\xi^{k}=\partial\xi_{1}^{k_{1}}\partial\xi_{2}^{k_{2}}...\partial\xi_{n}^{k_{n}}$. The complex variable characteristic function of operator $\hat{Q}(\xi,\eta)$ is
  \begin{eqnarray}\label{we9}
   && \chi_{Q}(\mu,\xi,\eta)=\chi_{Q}(0,\xi,\eta)\exp[-\frac{1}{2}(\mu,\mu^*)\tilde{\gamma}_{G}(\mu,\mu^*)^T\nonumber \\
   && -(\xi,\xi^*)\tilde{\gamma}_{G_{+}}(\mu,\mu^*)^T-(\eta,\eta^*)\tilde{\gamma}_{G_{-}}(\mu,\mu^*)^T],
  \end{eqnarray}
  where $\tilde{\gamma}_{G_{\pm}}=\tilde{\gamma}_{G}\pm \sigma_{1}\otimes I_{n}$ with the first Pauli matrix $\sigma_{1}$ and the $n\times n$ identity matrix $I_{n}$, and
  \begin{eqnarray}
    &&\chi_{Q}(0,\xi,\eta)=\exp[-\frac{1}{2}(\xi,\xi^*)\tilde{\gamma}_{G_{+}}(\xi,\xi^*)^T \nonumber\\
    && -(\xi,\xi^*)\tilde{\gamma}_{G_{-}}(\eta,\eta^*)^T-\frac{1}{2}(\eta,\eta^*)\tilde{\gamma}_{G_{-}}(\eta,\eta^*)^T].
  \end{eqnarray}
  We thus have
  \begin{eqnarray}\label{we10}
     && Tr(\hat{\rho} \hat{M})=\mathcal{N}\mathcal{O}Tr[\hat{Q}(\xi,\eta)\hat{M}]\nonumber\\
     &&=\mathcal{N}\mathcal{O}\int[\frac{d^2\mu}{\pi}]\chi_{Q}(\mu,\xi,\eta)\chi_{M}(\mu)\nonumber\\
     && =\frac{\mathcal{N}\mathcal{O}\chi_{Q}(0,\xi,\eta)\exp[f(\xi,\eta)]}{\sqrt{|\det(\tilde{\gamma}_{G}+\tilde{\gamma}_{M})|}}
  \end{eqnarray}
  with $f(\xi,\eta)=\frac{1}{2}[(\xi,\xi^*)\tilde{\gamma}_{G_{+}}+(\eta,\eta^*)\tilde{\gamma}_{G_{-}}](\tilde{\gamma}_{G}+\tilde{\gamma}_{M})^{-1}[\tilde{\gamma}_{G_{+}}(\xi,\xi^*)^T+\tilde{\gamma}_{G_{-}}(\eta,\eta^*)^T]$. In the limit of $\tilde{\gamma}_{M}\rightarrow\infty$ (also denoted as $\gamma_{M}\rightarrow\infty$), that is, all the parameters, $M_{i}$, in the $\gamma_{M}$ tend to infinite or negative infinite, we have $f(\xi,\eta)\rightarrow 0$. Further notice that $Tr(\hat{\rho})=\chi_{\hat{\rho}}(0)=\mathcal{N}\mathcal{O}\chi_{Q}(0,\xi,\eta)=1$. Then
  \begin{equation}\label{we11}
    Tr(\hat{\rho} \hat{M})_{\gamma_{M}\rightarrow\infty}\rightarrow \frac{1}{\sqrt{|\det(\tilde{\gamma}_{G}+\tilde{\gamma}_{M})|}}=\frac{1}{\sqrt{|\det(\gamma_{G}+\gamma_{M})|}}.
  \end{equation}

  \begin{proposition}\label{proposition3}
  The necessary and sufficient criterion of separability for a photon added (subtracted) Gaussian state  of $1\times1$ system and some $2\times2$ system is
  \begin{equation}\label{we12}
   \min_{\gamma_{M}}\max_{\gamma_{A},\gamma_{B}}\frac{\det(\gamma_{G}+\gamma_{M})}{\det(\gamma_{A}\oplus\gamma_{B}+\gamma_{M})}\geq 1,
\end{equation}
for $\gamma_{M}$ with infinite parameters $M_{i}$. Herein, $\gamma_{G}$ is the CM of Gaussian kernel of photon added (subtracted) state $\hat{\rho}$, $\gamma_{G}$ is either a $1\times1$ CM, or a $2\times2$ CM in the form of (\ref{we3}), $\gamma_{A}$ and $\gamma_{B}$ are the CMs of the subsystem pure Gaussian states with $\det(\gamma_{A})=\det(\gamma_{B})=\frac{1}{4}$.
\end{proposition}

The necessary part of criterion (\ref{we12}) comes from (\ref{we11}) and $Tr(\hat{\rho} \hat{M})\leq \Lambda$ if $\hat{\rho}$ is separable, where $\Lambda=[\det{(\gamma_{A}\oplus\gamma_{B}+\gamma_{M})}]^{-\frac{1}{2}}$ has already been established for a Gaussian detector in $1\times1$  and some $2\times2$ CV systems.

The reason of sufficiency is as follows. The criterion (\ref{we12}) is sufficient for the separability of the kernel state $\hat{\rho}_{G}$, as indicated by (\ref{we2f}) and (\ref{we3e}). If the condition (\ref{we12}) is true, then $\hat{\rho}_{G}$ is separable, the state $\hat{\rho}$ is separable too since $\hat{\rho}$ is prepared from a separable state $\hat{\rho}_{G}$ with local operations (photon addition and/or subtraction ).

The criterion (\ref{we12}) means that the necessary and sufficient criterion of separability for a photon added (subtracted) state can be reduced to the necessary and sufficient criterion of separability of its Gaussian kernel under the condition of infinite $\gamma_{M}$. The application of (\ref{we12}) leads to the following corollary.

\begin{corollary}\label{c1}
For a $1\times1$ NGPASG state $\hat{\rho}=\mathcal{N}\hat{a}_{1}^{\dagger k_{1}}\hat{a}_{2}^{\dagger k_{2}}\hat{a}_{1}^{m_{1}}\hat{a}_{2}^{m_{2}}\hat{\rho}_{G}\hat{a}_{1}^{\dagger m_{1}}\hat{a}_{2}^{\dagger m_{2}}\hat{a}_{1}^{k_{1}}\hat{a}_{2}^{k_{2}}$ with Gaussian kernel described by stardard CM of (\ref{we2g}) and a $2\times2$ NGPASG state with Gaussian kernel characterized by CM of (\ref{we3}), the necessary criteria of separability are (\ref{we2l}) and (\ref{we3f}), respectively, regardless the number of photon added to or subtracted from the Gaussian kernel.
\end{corollary}

The special case of necessary and sufficient condition of separability for a two-mode NGPASG prepared by a single photon adding to (subtracting from) a symmetric Gaussian state at each mode has been shown in \cite{Jiang14}.

{\it{ Conclusion}} -
We use a Gaussian operator to build Gaussian witness for detecting the entanglement of either Gaussian states or non-Gaussian states prepared from Gaussian kernel states by photon additions or/and subtractions. Necessary and sufficient separable criteria are given for $1\times1$ Gaussian and related non-Gaussian states, generalized Werner-Wolf $2\times2$ Gaussian states and related non-Gaussian states, respectively. The validity of the Gaussian witness is to show that the maximal mean of Gaussian operator over product states is achieved by product of Gaussian pure states, they are vacuum, coherent states and squeezed states in the cases considered. We show the validity using the known properties of the Bosonic Gaussian channel of gauge covariant or gauge contracovariant. We transform the maximal mean problem to a Bosonic Gaussian channel problem when the covariance matrix of the Gaussian detector operator tends to infinite.

Further progresses could be made for multi-mode Gaussian witnesses as far as the corresponding Bosonic Gaussian channels are diagonalizable \cite{GiovannettiCM}.

{\it{Acknowledgement}}-This work is supported by the National Natural Science Foundation  of China (Grant No.61871347)

\end{document}